\documentclass[preprint,showpacs,showkeywords,preprintnumbers,amsmath,amssymb]{revtex4}
\usepackage{graphicx}
\usepackage{dcolumn}
\usepackage{bm}
\begin{document}

\title{Internal and External Fluctuation Activated Non-equilibrium Reactive Rate Process}

\author{WANG Chun-Yang}
\thanks{Corresponding author. Email: wchy@mail.bnu.edu.cn}

\affiliation{Department of Physics and engineering, Qufu Normal
University, Qufu, 273165, China}

%\date{\today}

\begin{abstract}
The activated rate process for non-equilibrium open systems is
studied taking into account both internal and external noise
fluctuations in a unified way. The probability of a particle
diffusing passing over the saddle point and the rate constant
together with the effective transmission coefficient are calculated
via the method of reactive flux. We find that the complexity of
internal noise is always harmful to the diffusion of particles.
However the external modulation may be beneficial to the rate
process.
\end{abstract}


\maketitle

\section{\label{sec:level1}INTRODUCTION}

More than seventy years ago H. A. Kramers published a seminal work
on the diffusion model of chemical reactions \cite{kramers}. Ever
since then, the theory of activated processes has become a central
issue in many fields of study \cite{aps1,aps2}, notably in chemical
physics, nonlinear optics and condensed matter physics. In the
model, the particle was supposed to be immersed in a huge
equilibrium medium so that it can gain enough energy to cross the
barrier from its thermally activation. The common feature of an
overwhelming majority of such treatments is that the system is
thermodynamically closed. That is to say, the noise of the medium is
of internal origin so that the dissipation-fluctuation theorem
\cite{fdt1,fdt2} is satisfied and a zero current steady state
situation is characterized by an equilibrium Boltzmann distribution.

However, when the system is thermodynamically open, for example,
driven by an external noise which is independent of the medium
\cite{whor}, no relation between the dissipation and fluctuations
can be dependent on. The corresponding situation aforesaid, if
attainable, can then only be defined by a steady state condition
\cite{dsray2,dsray3} which may depend not only on the strength and
correlation of the external noise but also on the dissipation of the
system. The external noise will modify the dynamics of activation in
the region around the barrier top so that an unusual effective
stationary flux across it gets resulted. This would in no doubt
induce an unfamiliar activated barrier escaping process that is
worth pursuing for the rate theory.

Therefore we present in this paper a recent study of us on the
activated rate process where the diffusing particle is under the
joint influence of the internal noise combining an external one. The
paper is organized as follows: In Sec. II, reactive dynamics at the
barrier top is investigated by analytically solving the generalized
Langevin equation. In Sec. III, we give a detailed discussion about
the combined effect of internal and external noises on the rate
process by asymptotically calculating the rate function and its
transmission coefficient. Sec. IV serves as a summary of our
conclusion where some implicate applications of this study are also
discussed.

\section{Reactive dynamics at the barrier top}

We consider the motion of a particle of unit mass moving in a
Kramers type potential $U(x)$ such that it is acted upon by random
forces $\zeta(t)$ and $\epsilon(t)$ of both internal and external
origin, respectively, in terms of the following generalized Langevin
equation (GLE):
\begin{eqnarray}
\ddot{x}+\int^{t}_{0}dt'\gamma(t-t')\dot{x}(t')+\partial_{x}U(x)
=\zeta(t)+\epsilon(t),\label{eq.GLE}
\end{eqnarray}
where $U(x)$ is the potential, the friction kernel $\gamma(t)$ is
connected to internal noise by the well-known
fluctuation-dissipation theorem (FDT) \cite{fdt1,fdt2}
$\langle\zeta(t)\zeta(t')\rangle=k_{B}T\gamma(t-t')$. Both the
noises $\zeta(t)$ and $\epsilon(t)$ are assumed stationary and
Gaussian with arbitrary decaying type of correlation. We further
assume, without any loss of generality, that $\zeta(t)$ is
independent of $\epsilon(t)$ so that we have
$\langle\zeta(t)\epsilon(t')\rangle=0$ and
$\langle\epsilon(t)\epsilon(t')\rangle_{e}=2D\psi(t-t')$. Here
$\langle\cdots\rangle_{e}$ implies the averaging over all the
realizations of $\epsilon(t)$ with $D$ the intensity constant and
$\psi(t)$ a relevant memory function. In other words, the external
noise is independent of the friction kernel $\gamma(t)$ and so there
is no corresponding fluctuation-dissipation relation. However,
correlation $\langle\epsilon(t)\epsilon(t')\rangle_{e}=2D\psi(t-t')$
is reminiscent of the familiar FDT formula due to the appearance of
the external noise intensity, it serves rather as a thermodynamic
consistency condition instead.

Due to the Gaussian property of the noises $\zeta(t)$ and
$\epsilon(t)$ and the linearity of the GLE, the joint probability
density function of the system oscillator must still be written in a
Gaussian form \cite{adelm} The reduced distribution function can
then be yielded by integrating out all the variants except $x$ as
\begin{eqnarray}
W(x,t; x_{0},v_{0}) =\frac{1}{\sqrt{2\pi
}\sigma_{x}(t)}\textrm{exp}\left[{-\frac{(x-\langle
x(t)\rangle)^{2}}{2\sigma^{2}_{x}(t)}}\right].\label{eq,pd}
\end{eqnarray}
in which the average position $\langle x(t)\rangle$ and variance
$\sigma^{2}_{x}(t)$ can be obtained by Laplace solving the GLE. In
the case of an inverse harmonic potential
$U(x)=-\frac{1}{2}m\omega^{2}_{b}x^{2}$, it reads
\begin{subequations}\begin{eqnarray}
\langle
x(t)\rangle&=&\left[1+\omega^{2}_{b}\int^{t}_{0}H(t')dt'\right]x_{0}+H(t)v_{0}\\
\sigma^{2}_{x}(t)&=&\int^{t}_{0}dt_{1}H(t-t_{1})\int^{t_{1}}_{0}dt_{2}\langle
\xi(t_{1})\xi(t_{2})\rangle H(t-t_{2})
\end{eqnarray}\label{aver}\end{subequations}
where $H(t)$ namely the response function can be yielded from
inverse Laplace transforming
$\hat{H}(s)=[s^{2}+s\hat{\gamma}(s)-\omega^{2}_{b}]^{-1}$ with
residue theorem \cite{ret1,ret2}. $\xi(t)$
$(=\zeta(t)+\varepsilon(t))$ is an effective noise of zero mean
$\langle\xi(t)\rangle=0$ whose correlation is given by
\begin{eqnarray}
\langle\xi(t)\xi(t')\rangle=\langle\zeta(t)\zeta(t')
\rangle+\langle\epsilon(t)\epsilon(t')\rangle_{e},\label{eq.cor}
\end{eqnarray}
where the two averages in the right hand side are taken
independently.

The probability of passing over the saddle point, namely also the
characteristic function which is crucial for the activated barrier
crossing process, can then be determined mathematically by
integrating Eq.(\ref{eq,pd}) over $x$ from zero to infinity as
\begin{eqnarray}
P(x_{0},v_{0};t)&=&\int^{\infty}_{0}W(x,t;x_{0},v_{0})dx,\nonumber\\
&=&\frac{1}{2}\textrm{erfc}\left[-\frac{\langle
x(t)\rangle}{\sqrt{2}\sigma_{x}(t)}\right],\label{eq,chsi}
\end{eqnarray}
The escape rate of a particle, defined in the spirit of reactive
flux method by assuming the initial conditions to be at the top of
the barrier, can then be yielded from
%\begin{widetext}
\begin{eqnarray}
k(t)&=&\frac{1}{h}\int^{\infty}_{-\infty}dx_{0}
\int^{\infty}_{-\infty}v_{0} W_{\textrm{st}}(x_{0},v_{0})P(
x_{0},v_{0};t)\delta(x_{0}-x_{b})dv_{0}\label{eq,rate}
\end{eqnarray}
%\end{widetext}
in the phase space. This in proceeding results in a generalized
transition state (TST) rate \cite{TST1,TST2,TST3}:
$k^{\textrm{TST}}=\frac{1}{Qh}e^{-U_{b}/(D_{b}+\Psi(\infty))}$ and
an effective transmission factor
\begin{eqnarray}
\kappa(t)&=&\left(1+\frac{\sigma^{2}_{x}(t)}{D_{b}H^{2}(t)}\right)^{-1/2},\label{eq,kappa}
\end{eqnarray}
where
$W_{\textrm{st}}(x_{0},v_{_{0}})=\frac{1}{Q}\textrm{exp}[-\{{\frac{v_{0}^{2}}{2D_{b}}+\frac{\tilde{U}(x_{0})}{D_{b}+\Psi(\infty)}}\}]$
is a Boltzmann form stationary probability distribution which can be
obtained from the steady state Fokker-Planck equation \cite{dsray3}.
This stationary distribution for the non-equilibrium open system is
not an equilibrium distribution but it plays the role of an
equilibrium distribution of the closed system, which may, however,
be recovered in the absence of the external noise.
$(D_{b}+\Psi(\infty))/k_{B}$ in the exponential factor of
$k^{\textrm{TST}}$ defines a new effective temperature
characteristic of the steady state of the non-equilibrium open
system.

In the particular case we have considered, parameters used
heretofore are defined as \cite{dsray2,dsray3}:
$\tilde{U}(x)=U_{b}-\frac{1}{2}\Omega_{b}^{2}(x-x_{b})^{2}$ the
renormalized linear potential near the barrier top with $\Omega_{b}$
an effective frequency and $U_{b}$ the barrier height.
$\Psi(\infty)$ and $D_{b}=\Phi(\infty)/\Gamma(\infty)$ are to be
calculated from
\begin{subequations}\begin{eqnarray}
\Phi(t)&=&\Omega_{b}^{2}(t)\sigma^{2}_{xv}(t)+\Gamma(t)\sigma^{2}_{v}(t)+\frac{1}{2}\frac{d}{dt}\sigma^{2}_{xv}(t),\\
\Psi(t)&=&\frac{d}{dt}\sigma^{2}_{xv}(t)+\Gamma(t)\sigma^{2}_{xv}(t)+\Omega_{b}^{2}(t)\sigma^{2}_{x}(t)-\sigma^{2}_{v}(t),
\end{eqnarray}\label{pp}\end{subequations}
for the steady state in which
$\Gamma(t)=-\frac{d}{dt}\textrm{ln}\Lambda(t)$,
$\Omega_{b}^{2}(t)=\dot{H}(t)(\dot{H}(t)-H(t))/\Lambda(t)$ and
$\Lambda(t)=\frac{\dot{H}(t)}{\omega_{b}^{2}}[1-\omega_{b}^{2}\int_{0}^{t}d\tau
H(\tau)]+H^{2}(t)$. Other variances besides $\sigma^{2}_{x}(t)$ are
also to be got from Laplace solving the GLE. These variables will
play a decisive role in the calculation of barrier escaping rate.
Therefore, in general, one has to work out these quantities first
for analytically tractable models \cite{adelm}.

As is expected, all the parameters besides the rates $\kappa(t)$ and
$k(t)$ are closely related to the internal and external noises.
Therefore a combining control of internal and external noise on the
activated barrier escaping process is prospected. This is what will
be involved in the following sections.

\section{Internal vs external noise}

Before accomplishing the following calculations, let us firstly
digress a little bit about $P(x_{0},v_{0};t)$ and $\kappa(t)$ which
are the central results of this study. As has been shown in Eqs.
(\ref{eq,chsi}) and (\ref{eq,kappa}), both the expressions of
$P(x_{0},v_{0};t)$ and $\kappa(t)$ are reminiscent of the familiar
previous results \cite{jcp,jdb}. Although variance
$\sigma^{2}_{x}(t)$ has been changed intrinsically by the external
noise, difference lives only superficially in the emergence of
$D_{b}$ which is an asymptotical constant in the long time limit.
Due to the independence of internal and external noises, other
variances such as $\langle x(t)\rangle$ and $H(t)$ depend only on
the internal noise. Therefore, from the viewpoint of diffusing
passing over the saddle point, the mean position of the Gaussian
packet relies simply on the internal noise while the width of it
depends not only on the internal noise but also on the external one.
It is the combining effect of internal and external noise that
determines the final diffusing process. In what follows, it shall be
concerned with several limiting situations to illustrate the general
result systematically for both thermal and non-thermal activated
processes.

\subsection{Internal white noise}\label{3a}

Firstly we consider the simplest case of a $\delta$ correlated
internal thermal noise combining with no external ones. To this end,
it is to set
\begin{eqnarray}
\epsilon(t)=0\hspace{0.5cm} \textrm{and} \hspace{0.5cm}
\langle\zeta(t)\zeta(t')\rangle=k_{B}T\gamma\delta(t-t').
\end{eqnarray}
By combining with the abbreviations in Eqs.(\ref{aver}) and
(\ref{pp}), all the quantities we need for the activated process can
be obtained easily. After some algebra it follows that
\begin{eqnarray}
\Psi(\infty)=0\hspace{0.5cm} \textrm{and} \hspace{0.5cm}
D_{b}=k_{B}T.
\end{eqnarray}
The relations obtained heretofore reduce to the general form
\begin{eqnarray}
\kappa(t)&=&\left(1+\frac{\sigma^{2}_{x}(t)}{k_{B}TH^{2}(t)}\right)^{-1/2}.\label{eq,kappa2}
\end{eqnarray}
This is a trivial result for the one-dimensional time-dependent
barrier passage \cite{jdb}. It generally describes the possibility
of a particle already escaped from the metastable well to recross
the barrier. In the case of no external noise modulation, it keeps
its usual form just as it should be.

\subsection{Internal color noise}

Next we discuss the case of purely internal color noise. For
example, we set the internal noise to be Ornstein-Uhlenbeck (OU)
type \cite{ou1,ou2}. That is
\begin{eqnarray}
\epsilon(t)=0\hspace{0.5cm} \textrm{and} \hspace{0.5cm}
\langle\zeta(t)\zeta(t')\rangle=k_{B}T\frac{G}{\tau_{c}}e^{-|t-t'|/\tau_{c}},
\end{eqnarray}
where $G$ denotes the strength while $\tau_{c}$ refers to the
correlation time of the noise. It should be noted that for
$\tau_{c}\rightarrow0$ the internal noise shown above becomes also
$\delta$ correlated. After some algebra we can obtain from
Eqs.(\ref{aver}) and (\ref{pp}) again that
\begin{eqnarray}
\Psi(\infty)=k_{B}T\left(\frac{\Omega_{b}^{2}}{\omega_{b}^{2}}-1\right)\hspace{0.5cm}
\textrm{while} \hspace{0.5cm} D_{b}=k_{B}T.
\end{eqnarray}
This will result in a different form of TST rate $k^{\textrm{TST}}$
but has no influence on the form of transmission coefficient
$\kappa(t)$. However, since in most cases
$\Omega_{b}^{2}\cong\omega_{b}^{2}$, the value of $\Psi(\infty)$ is
actually close to 0. The values of variance $\sigma^{2}_{x}(t)$ and
$H(t)$ are also changed intrinsically. Therefore, although the form
of $\kappa(t)$ has not been changed, the rate process has been
modified implicitly due to the alteration of internal noise.

\subsection{Internal and external white noise}

In further, let us turn to investigate the more complicated
combining case where both the internal and external noise to be
$\delta$ correlated, i.e.
\begin{eqnarray}
\langle\epsilon(t)\epsilon(t')\rangle_{e}=2\mathcal{D}\delta(t-t')\hspace{0.5cm}
\textrm{and}
\hspace{0.5cm}\langle\zeta(t)\zeta(t')\rangle=k_{B}T\gamma\delta(t-t').
\end{eqnarray}
in which $\mathcal{D}$ is the strength of the external white noise.
This may be the simplest combining case of internal and external
noise. Derive again from the abbreviations heretofore we obtain
\begin{eqnarray}
\Psi(\infty)=0 \hspace{0.5cm} \textrm{and} \hspace{0.5cm}
D_{b}=k_{B}T+\frac{\mathcal{D}}{\gamma}.
\end{eqnarray}
Noticing that comparing with the aforesaid case in Sec.\ref{3a} a
new effective temperature due to external noise is defined here by
$\mathcal{D}/(\gamma k_{B})$ \cite{eft}. But in the limit of
$\mathcal{D}\rightarrow0$, the previous one-dimensional form of
$\kappa(t)$ for pure internal white noise can still be recovered by
Eq.(\ref{eq,kappa}). However, the rate process has also been changed
intrinsically due to the effect of external noise.

\subsection{Internal color and external white noise}

Finally, we consider a particular case where the external noise is
$\delta$ correlated while the internal is an OU process, i.e.,
\begin{eqnarray}
\langle\epsilon(t)\epsilon(t')\rangle_{e}=2\mathcal{D}\delta(t-t')\hspace{0.5cm}
\textrm{and}\hspace{0.5cm}
\langle\zeta(t)\zeta(t')\rangle=k_{B}T\frac{G}{\tau_{c}}e^{-|t-t'|/\tau},\label{pt}
\end{eqnarray}
both symmetric with respect to the time argument and assumed to be
uncorrelated with each other. By virtue of similar derivations as
heretofore, we find it is difficult to get an explicitly simple
expression of $\Psi(\infty)$ while $D_{b}=k_{B}T$ is recovered
again. This is nontrivial in the presence of external noise because
it seems as if there is not any effect on the rate process that
comes from the external noise. But actually all the effects have
been contained in the calculations of $\sigma^{2}_{x}(t)\rangle$ and
$H(t)$ so as in that of $\kappa(t)$.

\begin{figure}[t]
%\centering
\includegraphics[scale=0.7]{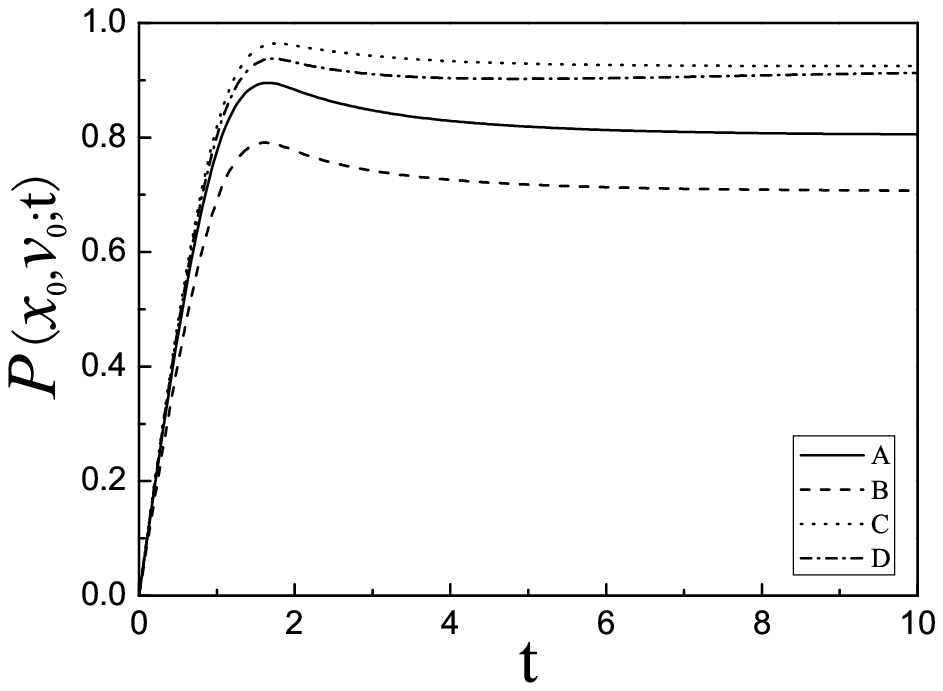}
\includegraphics[scale=0.7]{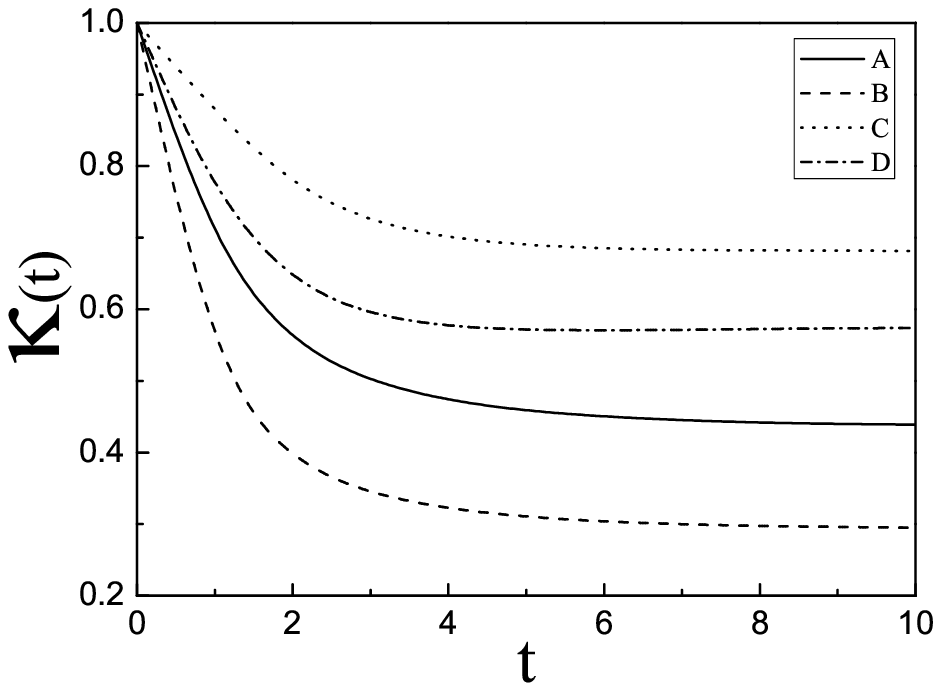}
\caption{Instantaneous values of $\chi(x_{0},v_{0};t)$ and
$\kappa(t)$ for various types of internal and external noise
combinations. Dimensionless parameters used are $\gamma=4.0$,
$\omega_{b}=1.0$, $G=3.0$ and $\tau=0.3$. Particles are assumed to
start from position $x_{0}=-1.0$ with initial velocity $v_{0}=6.0$.
Lines signed as $A$, $B$, $C$, $D$ represent the different cases
discussed in the corresponding subsections
respectively.}\label{Fig1}
\end{figure}

In order to give an explicit revelation of the combining effect of
the noises on the rate process, we plot in Fig.\ref{Fig1} the
instantaneous values of $P(x_{0},v_{0};t)$ and $\kappa(t)$ at
different types of combining cases aforesaid. From which we can see
that the asymptotic stationary value of $P(x_{0},v_{0};t)$ and
$\kappa(t)$ (defined as $P_{\textrm{st}}$ and $\kappa_{\textrm{st}}$
respectively) in the internal color case is smaller than that of
internal white case no matter the system is modulated by an external
white noise or not. On the contrary, both $P_{\textrm{st}}$ and
$\kappa_{\textrm{st}}$ are larger than those of pure internal case
supposing an external noise is set on with modulation. Thus we can
infer from considering the intrinsic meaning of $P(x_{0},v_{0};t)$
and $\kappa(t)$ that the complexity of internal noise (or
dissipation) is always harmful to the diffusion of particles.
However the external modulation may be beneficial to the rate
process. This is a non-trivial result of great meaning to many
different kinds of realistic situations in forming a non-equilibrium
(or non-thermal) system-reservoir coupling environment
\cite{phys1,phys2}.

\section{Summary and discussion}

In summary, we have studied in this paper the activated rate process
for non-equilibrium open systems taking into account both internal
and external noise fluctuations in a unified way. We calculated the
probability of a particle diffusing passing over the saddle point
and the rate constant together with the effective transmission
coefficient via the method of reactive flux. The combining control
of internal and external noises on the activated barrier escaping
process is investigated. We find that the complexity of internal
noise is always harmful to the diffusion of particles. However the
external modulation may be beneficial to the rate process.

We believe that these considerations are likely to be important in
other related issues in non-equilibrium open systems and may serve
as a basis for studying processes occurring within irreversibly
driven environments \cite{jray,rher} and for thermal ratchet
problems \cite{rdas}. The externally generated non-equilibrium
fluctuations can bias the Brownian motion of a particle in an
anisotropic medium and may also be used for designing molecular
motors and pumps.

\section * {ACKNOWLEDGEMENTS}

This work was supported by the Shandong Province Science Foundation
for Youths (Grant No.ZR2011AQ016) and the  Shandong Province
Postdoctoral Innovation Program Foundation (Grant No.201002015).

\end{document}